# COLD ATOMS AND CREATION OF NEW STATES OF MATTER: BOSE-EINSTEIN CONDENSATES, KAPITZA STATES, AND '2D MAGNETIC HYDROGEN ATOMS'

LENE VESTERGAARD HAU, B. D. BUSCH, CHIEN LIU, MICHAEL M. BURNS, AND J. A. GOLOVCHENKO

*Rowland Institute for Science, 100 Edwin H. Land Boulevard,
Cambridge, MA 02142, USA*

We have succeded in creating Bose-Einstein condensates with 2 million sodium atoms in a '4D' magnetic trap. We show the dynamic formation of a condensate as evaporative cooling proceeds. We also present a series of trap-release pictures clearly showing the distinctly different modes of expansion of condensate and thermal cloud. We further give two examples of wave guides for atomic de Broglie matter waves. One structure, the Kapitza wave guide, uses the interaction between an electrically polarizable atom and a charged wire. For stably bound orbits, a dynamical stabilization with time dependent potentials is necessary. This system, which can be tuned freely between classical and quantum regimes, shows chaotic behavior in the classical limit. The static counterpart of this wave guide leads to the introduction of the 'angular momentum quantum ladder'. The second wave guide structure is based upon the interaction between a current carrying wire and the magnetic dipole moment of an atom. A hydrogenic spectrum of bound states is derived through the concept of supersymmetry.

## 1 Introduction

Recently, methods have been developed to cool atoms to micro and even nano Kelvin temperatures [1-2]. This leads to the formation of completely new states of matter, Bose-Einstein condensates [3], Kapitza states [4], and '2D magnetic hydrogen atoms' [5]. The first part of the paper describes our recent success in creating a condenste of sodium atoms in our newly designed '4D' trap [6]. The availability of very cold atoms also introduces requirements for finding efficient means of transporting cold atoms, characterized by long de Broglie wave lengths, from one part of an experiment to another, typically quieter location for high precision, low-noise measurements. As is well known, a special wave guide is required for microwave propagation. Likewise, atomic matter waves require very special guiding structures and we shall discuss here two such structures based upon the interaction between polarizable atoms and thin metallic wires. This leads to the introduction of 'Kapitza states', '2D magnetic hydrogen atoms', and 'quantum ladders'.

## 2    Bose Einstein Condensation of Sodium in the '4D' Magnetic Trap

We recently succeeded in creating a Bose-Einstein condensate through a combination of laser [1] and evaporative cooling [2] of sodium atoms. Our laser cooling system includes a dark-spot version of the magneto-optic trap [7] where we collect $2 \cdot 10^9$ atoms at densities of $3 \cdot 10^{11}/cm^3$ in a few seconds. In the magneto-optic trap, we use the Doppler effect to viscously damp the atom motion in a configuration of three pairs of orthogonal counterpropagating laser beams tuned 20 MHz below the atomic resonance line corresponding to the F=2-3 hyperfine trasition within the D2 manifold. Atoms are loaded from a candlestick atomic beam source [8] which, through capillary action, provides for recirculation of uncollimated sodium atoms. To increase the efficiency of loading from the source into the magneto-optic trap we use a zero-crossing Zeeman slower [9] where radiation pressure from an intense laser beam is used to decellerate the atoms. As the atoms decellerate, the Doppler shift will change, and a spatially varying B field keeps the atoms resonant with the laser beam. After atoms are collected in the magneto-optic trap, they are cooled in polarization gradients [10] for a few ms to decrease the temperature further from 1 mK to 50 microkelvin. We finally optically pump the atoms into the F=1 hyperfine level. At this point all laser beams are turned off and a trapping B field is turned on. We use our new '4D' configuration which is described in detail in [6]. The trap is a layered version of the Ioffe trap [11]. The generic Ioffe design is a cylindrically symmetric trap with linearly varying B field components perpendicular to the symmetry axis and a parabolic variation of the field component along the axis. Our design gives large linear gradients (217G/cm) and curvature (124 $G/cm^2$) for relatively low electric power consumption (7.2 kW). The layered structure allows the laser beams to enter the trapping region.

After the atoms are loaded into the trap, we evaporatively cool the cloud for 38 secs. In the evaporative cooling process the hottest atoms are kicked out by spin flipping them to states repelled by the magnetic field. We apply a transverse magnetic RF field tuned to the spin precession frequency corresponding to the largest B field seen by the hottest atoms [12]. The remaining atoms will collide and reequilibrate at a lower temperature. Subsequently the RF frequency is lowered and the next set of atoms is kicked out. We continue in this manner and end up with a cloud of 2 million atoms cooled below the transition temperature for Bose-Einstein condensation. We can control the final temperature and density of the atoms by adjusting the final RF cut frequency.

We observe the final atom cloud by imaging the transmission profile of a resonant laser probe beam right after the cloud onto a CCD camera. In Fig. 1 we show a series of absorption pictures taken after the trapping B field is turned off and the atoms are allowed to expand for 8 ms. The pictures are taken for varying final RF cut frequency where (a) shows a purely thermal cloud with a classical Gaussian density distribution. The condensate shows up in (b) where a two-component density distribution is clearly seen. As we cut still deeper, the condensate fraction grows - we see more condensate atoms at the expensse of atoms in the thermal cloud. In the last picture we end up with a pure condensate.

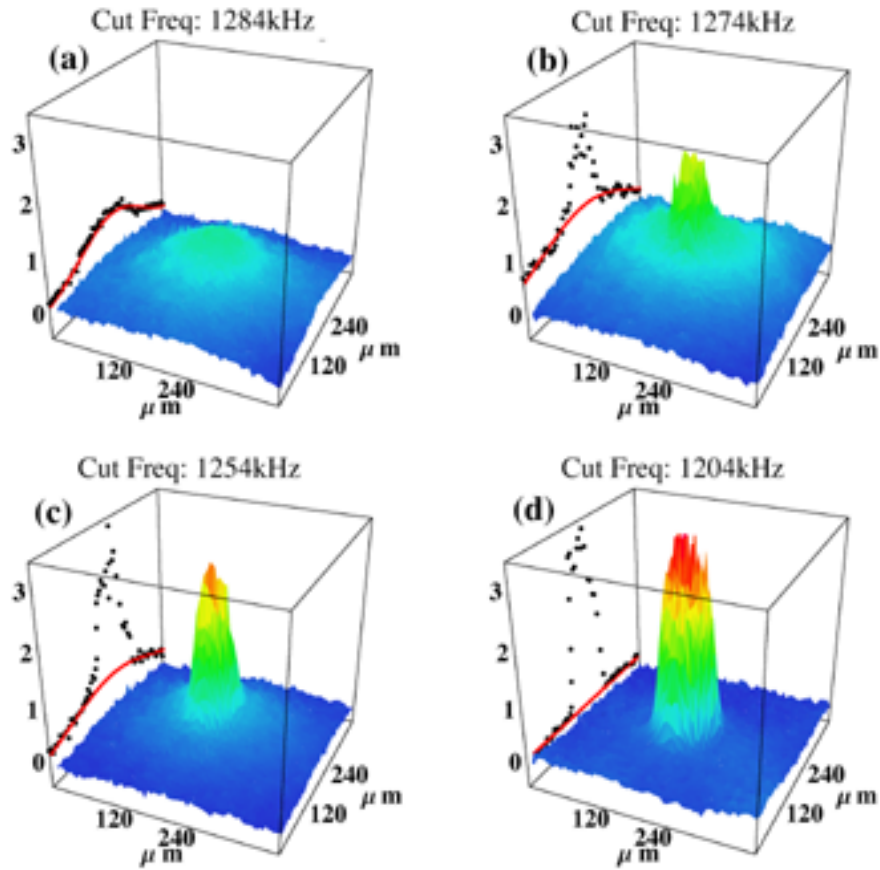

Figure 1: Formation of a Bose-Einstein condensate. The final RF cut frequency is noted on the figure in each case. (a) shows a purely thermal atom cloud. In (b) we have crossed the critical temperature for Bose-Einstein condensation at 530 nK. In (c) the condensate fraction has increased and in (d) we are left with an almost pure condensate: the condensate fraction is 70 % and the number of atoms in the condensate is 500,000.

In Fig. 2 is seen a series of pictures showing the expansion of the atom cloud as a function of time after release of the magnetic trap. The condensate expands highly anisotropically due to a large repulsive interaction between condensate atoms. In [13], we analyze the condensate expansion in detail and compare to calculations based on a time-dependent mean-field equation.

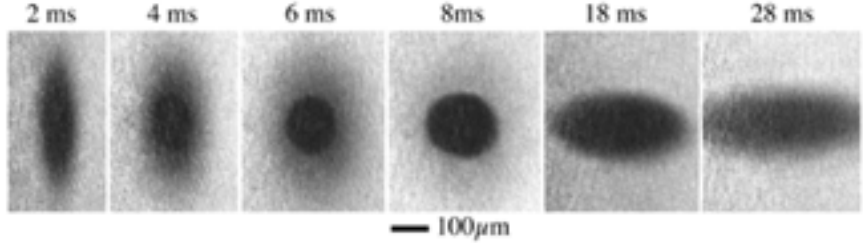

Figure 2: Expansion of atom clouds as a function of time after release of the magnetic trap. Condensates containing $10^6$ atoms are seen to expand highly anisotropically. The expansion rate of the thermal clouds indicates a temperature of 340 nK in the magnetic trap before release. The thermal cloud spreads outside the field of view of the camera after 10 ms.

## 3   Wave Guides for Atomic de Broglie Matter Waves

With the availability of methods to cool atoms to ultra low temperatures, we need a means to manipulate and transport long wavelength de Broglie matter waves. We shall present two such wave guides. The first is based upon bound atom motion around a charged wire.

The interaction between a neutral, electrically polarizable atom and the electric field from a charged wire is described by an attractive $1/r^2$ potential [4],

$$v(r) = -\alpha \frac{V^2}{2\ln^2\left(\frac{R}{r_0}\right) r^2} , \qquad (1)$$

where V is the electric voltage on the wire, $\alpha$ the atomic polarizability, $r_0$ the wire radius, and R the radius of an outer grounding cylinder (introduced to give the system a well defined capacity).

*3.1    Static $1/r^2$ Potential: The Angular Momentum Quantum Ladder*

The $1/r^2$ potential is at the border-line between regular and highly singular potentials and has very interesting properties: as seen from the uncertainty principle, the ground state in this potential collapses to the origin and the ground state energy is at minus infinity.

The positive energy states in this potential also have peculiar features. The radial motion of the particle is described by an effective potential, u(r), which is the sum of the potential in (1) and a centrifugal barrier term,

$$u(r) = \frac{L^2 - K^2}{2Mr^2}, \quad with \quad K^2 = \alpha\, M \frac{V^2}{\ln^2\left(\frac{R}{r_0}\right)}. \quad (2)$$

Here, L is the angular momentum along the direction of the wire for an atom with mass M. For an incident atomic beam, atoms with angular momentum below the critical value

$$L_{crit} = K \quad (3)$$

will experience a purely attractive potential and hit the wire. In the case of sodium atoms hitting the surface of a hot, oxygenated rhenium wire, for example, the atoms will ionize with almost 100% probability [14]. Thereby, the number of atoms in the beam with angular momentum below the critical value can be deduced from a measurement of the current in the wire. With a uniform distribution of angular momenta in the atomic beam, the current increases linearly with the voltage on the wire. Quantum mechanically, however, angular momentum is quantized and, as seen from a semiclassical description, the current will increase in steps, generating an 'angular momentum quantum ladder'.

With a full quantum mechanical description of this system, a smearing of the steps is expected from quantum tunneling to the wire. The Hamiltonian operator governing the radial part of the atomic motion outside the wire is given by

$$H_r = -\frac{\hbar^2}{2M}\frac{d^2}{dr^2} + u(r), \text{ where } L^2 = (l\hbar)^2 - \frac{1}{4}, \quad (4)$$

and $l$ is the angular momentum quantum number. Inelastic processes at the surface of the wire are taken into account by imposing appropriate boundary conditions on the radial wavefunction. The assumption of a "sticky surface" (the fact that atoms ionize with unit probability at the surface of the wire) is represented by introducing a step potential at the wire surface and assuming that the wavefunction has no outgoing radial current component just inside the surface. By assuming a continuous logarithmic derivative of the radial wavefunction, we match the purely ingoing radial current on the inside of the step to a combination of outgoing and ingoing radial current components of the wavefunction just outside the step. The absorption cross section is obtained by calculating the net ingoing current flux through a surface outside the wire. The step height is chosen such that in the limit of zero voltage, the cross section approaches the geometric size of the wire.

An example of a full quantum calculation is shown in Fig. 3 where the absorption cross section is shown as a function of voltage on the wire. The step heights $\Delta\sigma$ are on the order of the reduced de Broglie wavelength,

$$\Delta\sigma = 2\frac{\hbar}{Mv_\theta}, \tag{5}$$

where $v_\theta$ is the rotational velocity around the wire, exactly as would be obtained by a semiclassical analysis. For 10-20 micro Kelvin atoms, the steps are separated by a macroscopic voltage difference of 0.1 V (as obtained from Eqs 2-3), which is easily measurable. A precise measurement of the electric polarizability of atoms can be obtained from a measurement of the voltage distance between these steps.

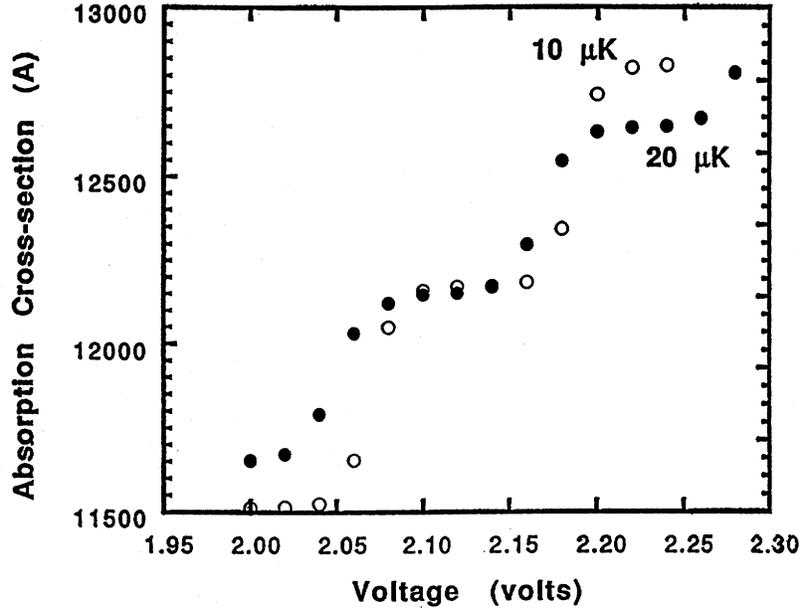

Figure 3: 'Angular Momentum Quantum Ladder'. The cross section for an atom to hit the wire as a function of the voltage on the wire. The figure is a result of a full quantum mechanical description where the ionization on the wire surface is taken into account through appropriate boundary conditions. The open circles are for sodium atoms at a temperature of 10 microkelvin and the solid circles for atoms of 20 microkelvin. The wire diameter is 0.1 micron.

### 3.2    *Dynamical Stabilization: Kapitza States*

A wave guide based upon a static $1/r^2$ potential, as described above, would be a disaster: all bound states hit the wire  This is due to the fact that the interaction potential in Eq. (1) between atom and wire is the attractive counterpart of the centrifugal barrier. Hence, the latter is not divergent enough to dominate for the smallest distances, closest to the wire.

However, we have discovered [4] that it is possible to stabilize atomic motion around the wire through application of a time varying potential on the wire. With a sinusoidally varying potential, the radial motion of an atom is described by a sum of the static potential in Eq. (2), and a time periodic potential

$$v(r,t) = W(r)\cos \omega t. \qquad (6)$$

Fig. 4 shows the radial motion with (a) a static electric voltage and (b) a time periodic electric voltage on the wire, where the peak voltage approximately equals the static voltage in (a). Similar initial conditions are used in the two cases and the figure is generated from a numerical solution to the classical equation of motion. In (a), the particle hits the wire in a few microseconds, and in (b) the motion is clearly stabilized. In the latter case, the radial motion separates into a fast and a slow motion where the fast part - the micromotion - is at a frequency given by the AC drive on the wire. The slow, secular motion is revealed by performing a time dependent canonical transformation to eliminate the micromotion from the equations of motion and a constant of motion is identified in this time dependent problem.

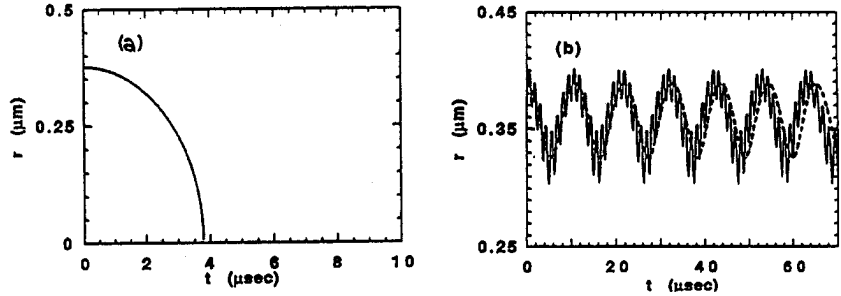

Figure 4: Radial motion for a sodium atom ($\alpha=27\text{Å}^3$) with angular momentum $L = 48\hbar$. Plot (a) corresponds to a static voltage of $8/\sqrt{2}$ volts on the wire and plot (b) to a time dependent voltage with a peak voltage of 8 volts and a frequency of 400 kHz. Similar initial conditions were applied in the two cases.

The canonical transformation generates an effective, repulsive potential in the Hamiltonian governing the slow motion. This potential, which we denote the

Kapitza potential [4] (after P. Kapitza who first described the stabilization of the inverted pendulum), is the time average of the kinetic energy in the micromotion,

$$V_{Kap} = \frac{(W')^2}{4M\omega^2} \quad . \tag{7}$$

In the wire case, the Kapitza potential is a highly divergent $1/r^6$ potential, which will repel the atoms from the wire and thus lead to stable, bound motion. The resulting total potential, which is a sum of the Kapitza potential and the static potential in Eq. (2), is shown in Fig. 5a. From the position of the potential minimum and the size of the well depth, we conclude that atoms cooled to microkelvin temperatures and submicron wires are required. A quantum mechanical description is called for and in Fig. 5b we show the quantized energy levels, corresponding to the bound 'Kapitza states', in the total effective potential as a function of angular momentum. The minimum of the potential (also shown) depends on angular momentum through the centrifugal barrier.

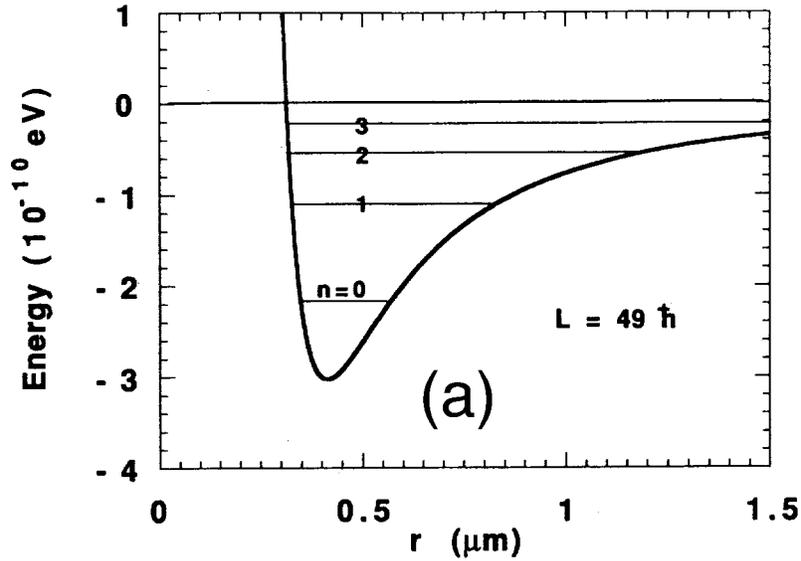

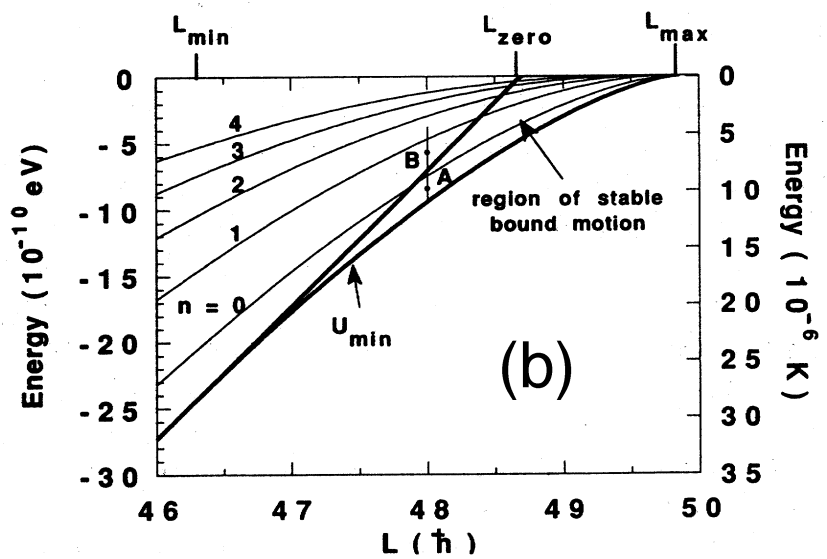

Figure 5: (a) Total effective potential (including the Kapitza potential) for the secular motion of sodium atoms with angular momentum $L = 49\hbar$ in the case of a time dependent voltage applied to the wire, where the peak voltage is 8 volts and the frequency 400 kHz. Also shown are the lowest quantized levels. (b) Quantized levels in the total effective potential for the secular motion as a function of angular momentum. The figure is obtained for the same time periodic voltage applied to the wire as in (a). The lower curve shows the minimum of the total effective potential and the upper curve (the line hitting the L axis in the point denoted $L_{zero}$) defines the border line of stability. Stable motion is obtained for points below the line.

The Kapitza description is obtained to lowest order in an expansion parameter that can be described as the ratio between the local oscillation frequency at the point of the atom orbit closest to the wire and the AC drive frequency. For higher excited states, the atom gets closer to the wire and the approximation eventually breaks down leading to unstable motion. The border line, separating stable from unstable parts of phase space, is shown in Fig. 5b: only states below this line will lead to stable motion.

      Larger peak values of the oscillating electric potential on the wire and a larger drive frequency lead to larger depths of the effective potential. With 800 Volts peak and a frequency of 4 MHz as one example we get a well depth - expressed in temperature units - on the order of millikelvin. This case results in a dense level spacing of bound states in the potential and is in this sense much more classical than the case represented by Fig. 5. It should be noted that the stabilizing, effective

Kapitza potential, created by a DC rectification of the AC potential, is responsible also for the fact that an inverted pendulum can be stabilized by dithering the point of support at a high frequency, for stability in the Paul trap for ions, and for the effect of strong focusing in accelerators.

We have generated Poincare plots for classical motion in the time dependent potential [4]. For initial conditions corresponding to the deepest bound levels in the total effective potential, we obtain closed tori in the Poincare plots. As the border line of stability is approached, islands of stability form. They correspond to resonance conditions where the ratios between drive and orbit frequencies are rational numbers. For higher energies yet, the stability is lost. We interpret the island formation as a sign of the onset of chaotic behavior.

The dynamical aspects of this system are thus very intriguing: it is possible to tune it freely between quantum and classical regimes and hence possible to study the quantum limit of a system that shows classical chaotic behavior.

*3.3     The '2D Magnetic Hydrogen Atom'*

An alternative wave guide for cold atoms is based upon the time-independent interaction of the atomic magnetic dipole moment and the magnetic field from a current carrying wire [5,16]. This system requires a quantum mechanical description in terms of multi component wave functions. We have found that the energy levels for bound states form a hydrogenic spectrum (exact for spin 1/2 particles and approximate for spin-one atoms). We have used the concept of supersymmetry, which we have extended to the case of multi component wavefunctions, to obtain the energy levels and eigen states for this '2D magnetic hydrogen atom' [5].

## 4     CONCLUSION

We have demonstrated the existence of new states of matter for atoms, laser and evaporatively cooled to micro and nano kelvin temparatures. We have created a Bose-Einstein condensate of sodium atoms in a '4D' magnetic trap. We have also fabricated thin metallic wires on which the two suggested types of wave guide for atomic de Broglie waves could be based. A particularly intriguing aspect of this research is the possibility to condense atoms directly into the bound states around a thin wire and furthermore to use the wire states as cavity states in a C.W. atom laser.

**ACKNOWLEDGEMENT**

This research was completed with support from the Rowland Institute for Science. L.V.H. in particular wants to thank Dr. Phil Dubois for his continued support and encouragement during the often frustrating periods of trying to 'get that condensate'.